\documentclass[sigconf]{acmart}
\AtBeginDocument{%
  }

\setcopyright{acmlicensed}
\copyrightyear{2026}
\acmYear{2026}
\setcopyright{cc}
\setcctype{by}
\acmConference[SIGIR '26]{Proceedings of the 49th International ACM SIGIR Conference on Research and Development in Information Retrieval}{July 20--24, 2026}{Melbourne, VIC, Australia}
\acmBooktitle{Proceedings of the 49th International ACM SIGIR Conference on Research and Development in Information Retrieval (SIGIR '26), July 20--24, 2026, Melbourne, VIC, Australia}
\acmDOI{10.1145/3805712.3808501}
\acmISBN{979-8-4007-2599-9/2026/07}

\usepackage{multirow}
\usepackage{soul}
\usepackage{xcolor}

\begin{document}

\title{From Unstructured to Structured: LLM-Guided Attribute Graphs for Entity Search and Ranking}

\author{Yilun Zhu}
\affiliation{%
  \institution{Amazon.com, Inc.}
  \city{Seattle}
  \country{USA}}
\email{yilunzhu@amazon.com}

\author{Nikhita Vedula}
\affiliation{%
  \institution{Amazon.com, Inc.}
  \city{Seattle}
  \country{USA}}
\email{veduln@amazon.com}

\author{Shervin Malmasi}
\affiliation{%
  \institution{Amazon.com, Inc.}
  \city{Seattle}
  \country{USA}}
\email{malmasi@amazon.com}

\renewcommand{\shortauthors}{Zhu et al.}

\begin{abstract}

Entity search, i.e., finding the most similar entities to a query entity, faces unique challenges in e-commerce, where product similarity varies across categories and contexts. Traditional embedding-based approaches often struggle to capture nuanced context-specific attribute relevance. In this paper, we present a two-stage approach combining Large Language Model (LLM)-driven attribute graph construction with graph-aware LLM ranking. In the offline stage, we extract structured product attributes from unstructured text, and construct a reusable attribute graph with category-aware schemas. In the online stage, we rank retrieved candidates by reasoning over this structured representation rather than raw text, reducing per-product token usage by 57\% while improving ranking precision. Experiments show that our approach outperforms multiple baselines under zero-shot scenarios, achieving a over 5\% improvement in average precision without requiring training data, generalizes robustly across diverse product categories, and shows immense potential for real-world deployment.

\end{abstract}

\ccsdesc[500]{Information systems~Language models}
\ccsdesc[300]{Information systems~Query representation}
\ccsdesc[500]{Computing methodologies~Information extraction}

\keywords{Entity Search, Structured Extraction, LLM Ranking}

\maketitle

\section{Introduction and Background}

\begin{figure}[t!]
    \centering
    \includegraphics[width=\columnwidth]{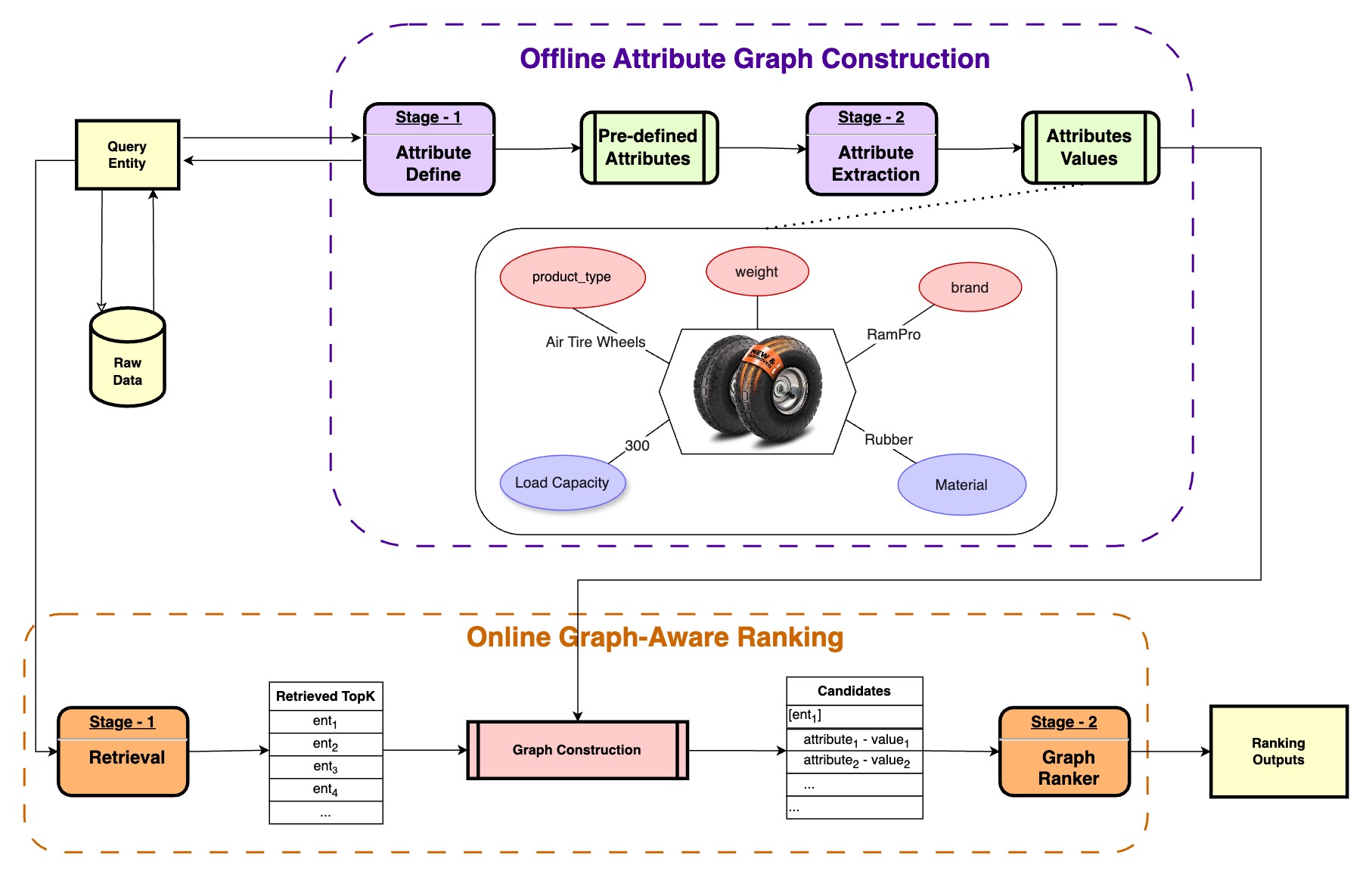}
    \Description{Two-stage architecture diagram: an offline stage constructs an attribute graph from unstructured product data using LLM-based extraction, and an online stage performs graph-aware LLM ranking of retrieved candidates.}
    \caption{The framework of our Graph Ranker System.}
    \label{fig:framework}
\end{figure}

\textit{Entity Search}, the task of retrieving entities most similar to a given query entity, is fundamental in information retrieval~\cite{balog2018entity}, particularly in e-commerce where users seek substitutes or comparable products~\cite{nigam2019semantic, reddy2022shopping}. Given a query product, a system must accurately retrieve and rank other entities that closely match the query's characteristics. Although product information typically exists as unstructured text (e.g., titles, descriptions), effective entity similarity search relies on structured attribute representations for systematic comparison \cite{metzger2013qbees, yang-etal-2024-eave}. A key challenge is extracting structured attributes and values from unstructured data, where the attribute set is dynamic and must adapt as new products and categories emerge \cite{etzioni2008open, karamanolakis-etal-2020-txtract}. 
Extracting structured product attributes from unstructured text (e.g., titles, descriptions) has been widely explored via sequence labeling, span-based extraction, and few-shot methods through graph-based inference and external knowledge augmentation~\cite{zhang2021queaco, deng2022jpave, li-etal-2023-attgen, loughnane-etal-2024-explicit, gong2024multilabel}. 
Traditional approaches in industry typically follow two stages, retrieval and ranking, which face further limitations. Collaborative filtering struggles with cold-start problems \cite{xv2023search}, text-based similarity may miss nuanced attribute differences \cite{van2016unsupervised}, and creating robust attribute-driven similarity measures remains challenging when item data is unstructured or inconsistent \cite{blume-etal-2023-generative}. Dense semantic embedding models and graph neural networks (GNNs) have advanced product-focused entity retrieval systems by capturing semantic relationships in large catalogs and addressing cold-start problems~\cite{balog2018entity, nigam2019semantic, chatterjee2021entity, jafarzadeh2022learning, zhu2025hint, dhole2025generative}. 

Recent advances in Large Language Models (LLMs) offer a promising avenue. LLMs can serve as powerful zero-shot list-wise rankers of search results \cite{sun-etal-2023-chatgpt, adeyemi-etal-2024-zero, zhu2024llmir, pradeep2023rankvicunazeroshotlistwisedocument}, and recent work integrates them with structured or semi-structured data (e.g., knowledge graphs and tables) for enhanced reasoning~\cite{kim-etal-2023-kg, sun2024thinkongraph, zhu2024llmir}. 
However, these approaches feed unstructured product descriptions directly to the LLM. At an industrial scale, this poses challenges with respect to cost, latency, efficiency, and the ability to enforce hard constraints like matching on critical attributes and nuanced values when applied naively over raw text. 

In this paper, we present an \textit{Attribute-based Graph Ranker}, a hybrid system for enhanced similarity search that addresses the above identified limitations through a two-stage design as shown in Figure~\ref{fig:framework}: (i) an \textit{offline} stage that uses LLM-based reasoning to extract structured attributes from unstructured product text and constructs a standardized attribute graph, and (ii) an \textit{online} stage that integrates high-recall retrieval with graph-aware LLM ranking, where the LLM reasons over curated attribute sets rather than free-text descriptions, maintaining efficiency and interpretability. We differ from prior work by decoupling knowledge construction from ranking. By reasoning over structured attributes, our system yields two benefits: (1) a 57\% reduction in per-product token usage, lowering inference cost and latency, and (2) explicit attribute-value comparisons enabling more precise similarity assessments than free-text reasoning alone. Our contributions are:

(i) \textbf{LLM-based Attribute Graph Construction}: An LLM-driven pipeline that defines category-aware attribute schemas, extracts structured attribute-value pairs from unstructured text, constructs a bipartite graph with product entities and attribute nodes linked by value relations, and enables precise attribute-based entity retrieval. 

(ii) \textbf{Graph-aware LLM Ranking}: An LLM reasons over the structured attributes at inference time, improving ranking precision and interpretability.

(iii) Large scale evaluations show significant precision gains over strong baselines, and strong potential for real-world deployment. %

\section{Methodology}

\subsection{Stage 1: Attribute Graph Construction}
The goal of this offline \textit{LLM-based Attribute Graph Construction} stage is to enrich and index the massive entity dataset so that the online second stage (Section \ref{sec:stage2}) can be fast and effective.\footnote{Since category labels are missing from the original dataset, we use an LLM-based approach to predict and standardize these labels to ensure products are compared within the same category.} This stage includes: (i) Attribute Definition, and (ii) Attribute Extraction. All LLM-based steps utilize Claude 3.5 Sonnet v2 \cite{claude3.5sonnet}, selected based on a preliminary evaluation of 500 examples in which it achieved high precision, while meeting our latency requirements.

\paragraph{Attribute Definition}
Different product categories have different characteristic attributes (e.g., ``screen size'' for TVs vs.\ ``material'' for clothing). To guide the extraction of attributes, we define an attribute schema for each product super-category and sub-category, and prompt an LLM to find key attributes relevant to products in that category. We do this at two levels: broad attributes that apply to the super-category (e.g., \textit{Brand}, \textit{Model}, \textit{Dimensions} for ``Electronics'') and specific attributes for the sub-category (e.g., \textit{Screen Size}, \textit{Storage Capacity}, \textit{Camera Resolution} for ``Smartphones''). Hierarchical organization ensures super-category attributes (e.g., \textit{Safety Standards}) remain consistent across sub-categories while sub-category attributes reflect domain-specific granularity. After standardization, we have 61 super-categories and 4,940 sub-categories, with approximately 8--10 attributes per super-category and 6--8 additional per sub-category. We curate these LLM-generated attribute lists through lightweight manual standardization via clustering and deduplication (e.g., merging ``Screen Size'' and ``Display Size''), rather than defining attribute schemas from scratch. For unseen categories, the LLM generates a new schema on-the-fly using the same prompting approach -- new sub-categories inherit super-category attributes while the LLM suggests category-specific ones. These predefined attribute sets ensure naming consistency and guide the extraction stage.

\paragraph{Attribute Value Extraction}
We then extract attribute values for each entity in the catalog. 
For each entity, we input the product's textual data (title, description, bullet points) to the LLM with an instruction to output a JSON of attribute-value pairs, using that product category's attribute definition list as a guideline. The LLM extracts values for predefined attributes, and we also allow it to identify additional product-specific attributes from the text.

\subsection{Stage 2: Graph-Aware LLM Ranking}
\label{sec:stage2}
During inference, our system receives a query entity $ENT_q$ as input and outputs a ranked list of similar entities $L = [ent'_1, ent'_2, \dots, ent'_k]$. The inference process leverages structured attributes previously extracted during the offline stage 1. If attributes for the query entity were not computed offline, the system dynamically invokes real-time attribute extraction to generate attribute-value pairs for $ENT_q$.\footnote{Newly extracted attributes are cached for efficiency in future queries.} The online stage includes: (i) candidate generation and (ii) graph-aware LLM ranking.

\paragraph{Candidate Generation}
Given the query entity $ENT_q$, we first retrieve an initial set of candidate entities from the complete catalog $D$. We employ dense retrieval using the embedding model \texttt{bge-m3}~\citep{chen-etal-2024-m3} to encode both query and candidate entities into dense embeddings based on concatenated product titles, descriptions, and bullet points. We use FAISS~\citep{johnson2019billion} for efficient approximate nearest-neighbor search, retrieving the top-$K_d$ candidate entities with the highest cosine similarity scores, forming the initial candidate set $C$.

\paragraph{Graph-aware LLM Ranking}
The final stage ranks candidate entities according to their similarity to the query entity, by reasoning over a bipartite graph \( G = (V_P, V_A, E) \), where \( V_P \) represents product (entity) nodes, \( V_A \) represents attribute nodes, and each edge \( e \in E \) connects a product node \( p \) to an attribute node \( a \) with edge label \( val(p,a) \) denoting the attribute value for that product.\footnote{We use attributes as nodes and values as edge labels for efficiency. With millions of products, representing each unique attribute value as a separate node would create an intractably large graph.} Given a query entity \( ENT_q \), ranking the candidates in the candidate set \( C \) involves implicitly traversing and reasoning over the local subgraph \( G_q = (V_{P_q}, V_{A_q}, E_q) \subseteq G \), where \( V_{P_q} = \{ENT_q\} \cup C \), and \( V_{A_q} \) contains all attributes associated with the entities in \( V_{P_q} \).
We formulate the ranking task as a zero-shot, list-wise reasoning problem, prompting Claude 3.5 v2 to compare each candidate's attribute values against the query entity's attribute values. The similarity score $S(ENT_q, ENT_c)$ between query entity $ENT_q$ and candidate entity $ENT_c \in C$ is computed implicitly by the LLM via assessing the edges connecting both entities to common attribute nodes:

{\scriptsize
\[
S(ENT_q, ENT_c) = \mathcal{F}_{\text{LLM}}\bigl(\{(a,\; val(ENT_q,a),\; val(ENT_c,a)) \mid a \in V_{A_q}\}\bigr)
\]}

where \(\mathcal{F}_{\text{LLM}}\) represents the LLM's reasoning function, which implicitly assigns importance to attributes and evaluates how well attribute values match or differ between the query and candidate entities. Candidates with stronger attribute alignment to the query entity receive higher similarity scores. We use a fine-grained 0--100 similarity scale\footnote{0 indicates complete dissimilarity and 100 represents perfect feature matching.} following established IR evaluation practices \cite{voorhees2000trec} and recent LLM evaluation frameworks \cite{zheng2023judging}, allowing for more nuanced differentiation of relevance levels than coarser scales. The final ranked list \( L = \text{sorted}\{[ent'_1, \dots, ent'_k]\}	 \) is produced by sorting candidates according to their similarity scores \(S(ENT_q, ENT_c)\). By using structured reasoning over this local attribute graph, our system achieves high-precision, interpretable ranking results suitable for industry applications demanding accuracy and transparency.

All LLM interactions in our pipeline use structured JSON schemas with predefined attribute fields and deterministic settings (temperature=0) for reproducibility. The input to the LLM at each stage consists exclusively of curated product catalog data — sanitized attribute-value pairs during ranking, and product text with a constrained attribute list during extraction. No user-generated free text reaches the LLM, and output is validated against the expected schema. This controlled pipeline significantly mitigates prompt-based adversarial risks. Our prompts were iteratively refined and optimized for the Claude LLM during development, and the structured input format constrains the output space and reduces sensitivity to prompt phrasing compared to free-text prompting. We also note that our evaluation covers general consumer product categories. For highly specialized domains (e.g., medical supplies), some domain-specific fine-tuning or expert-curated attribute schemas may be necessary to achieve comparable extraction quality.

\section{Experiments and Results}

\begin{table}[htb]
\centering\small
\begin{tabular}{l|ccccc}
\toprule
\textbf{Method} & \textbf{P@1} & \textbf{P@3} & \textbf{P@5} & \textbf{MRR} & \textbf{mAP} \\
\midrule
Sparse Retrieval (\textsc{sr}) & 40.48 & 37.30 & 31.90 & 50.87 & 50.43 \\
Dense Retrieval (\textsc{dr}) & 51.36 & 44.55 & 35.17 & 56.42 & 54.95 \\
\textsc{dr} + raw-ranker & 54.76 & 46.83 & 39.52 & 60.00 & 57.99 \\
\textsc{dr} + graph-ranker & \textbf{57.14} & \textbf{50.00} & \textbf{47.14} & \textbf{63.97} & \textbf{60.42} \\
\bottomrule
\end{tabular}
\caption{Human evaluation on 200 query-candidate pairs. \textsc{sr} and \textsc{dr} indicate sparse and dense retrieval respectively.}
\label{tab:human_eval}
\end{table}

\begin{table*}[t!]
\centering
\small
\setlength{\tabcolsep}{4pt}
\begin{tabular}{@{}l *{4}{c} *{6}{c}@{}}
\toprule
\multirow{2}{*}{Method} & \multicolumn{4}{c}{nDCG Metrics (\%)} & \multicolumn{6}{c}{Precision-based Metrics (\%)} \\
\cmidrule(r){2-5} \cmidrule(l){6-11}
& nDCG@1 & nDCG@3 & nDCG@5 & nDCG@10 & P@1 & P@3 & P@5 & P@10 & MRR & mAP \\
\midrule
\multicolumn{11}{l}{\textbf{Eval Threshold} $\geq$ \textbf{80}} \\
\cmidrule(l){1-11}
Sparse Retrieval (\textsc{sr}) & 58.58 & 55.80 & 54.75 & 56.35 & 43.86 & 33.42 & 27.29 & 18.71 & 50.18 & 42.01 \\
Dense Retrieval (\textsc{dr}) & 75.27 & 75.60 & 77.30 & 86.59 & 49.08 & 38.71 & 33.02 & 25.40 & 55.57 & 50.40 \\
\textsc{dr} + raw-ranker & 77.66 & 77.87 & 79.39 & 88.00 & 51.95 & 41.18 & 34.89 & 26.36 & 57.70 & 52.78 \\
\textsc{dr} + graph-ranker & \textbf{79.96} & \textbf{80.02} & \textbf{81.47} & \textbf{89.86} & \textbf{56.85} & \textbf{43.49} & \textbf{36.71} & \textbf{27.28} & \textbf{63.13} & \textbf{58.02} \\
\midrule
\multicolumn{11}{l}{\textbf{Eval Threshold} $\geq$ \textbf{50}} \\
\cmidrule(l){1-11}
Sparse Retrieval (\textsc{sr}) & \multicolumn{4}{c}{\multirow{4}{*}{}} & 58.71 & 49.41 & 43.05 & 32.19 & 66.10 & 50.63 \\
Dense Retrieval (\textsc{dr}) & \multicolumn{4}{c}{} & 72.73 & 65.54 & 61.31 & 54.40 & 78.59 & 72.11 \\
\textsc{dr} + raw-ranker & \multicolumn{4}{c}{} & 74.78 & 68.04 & 63.55 & 55.90 & 80.12 & 74.15 \\
\textsc{dr} + graph-ranker & \multicolumn{4}{c}{} & \textbf{78.12} & \textbf{69.39} & \textbf{64.66} & \textbf{56.90} & \textbf{83.53} & \textbf{77.35} \\
\bottomrule
\addlinespace[4pt]
\end{tabular}
\caption{Entity similarity search evaluated using Nova Pro LLM. The evaluation threshold represents the minimum similarity score required for a query-candidate pair to be considered relevant. \textsc{sr} and \textsc{dr} denote sparse and dense retrieval respectively.}
\label{tab:results}
\end{table*}

\noindent\textbf{Dataset and Experimental Setup:}
We evaluate our proposed approach on a large real-world dataset, Amazon Shopping Queries \cite{reddy2022shopping}, prioritizing practical considerations like realistic data scale and manual evaluation where automated ground truth is lacking. This dataset contains challenging search queries paired with products and relevance labels (Exact, Substitute, Complement, Irrelevant). While the dataset is originally for query-to-product relevance, we repurpose it for our entity-to-entity search task by treating products as query entities, and considering all other products as candidates. There are 469,898 unique products under the \texttt{US} locale. For evaluation, we curate a diverse subset of 8.7K query entities sampled across various product super-categories. Each query entity is associated with its corresponding title, description, and other relevant information, from which we extracted pertinent attributes during a preprocessing stage. Certain product categories were excluded from consideration such as adult products (due to ethical concerns); books, media and video games (insufficient products and limited useful context to extract meaningful attributes); and categories with fewer than 500 products (to ensure sufficient data representation).

We compare our approach against the following baselines:

(i) \noindent\textbf{Retriever-only.} We implement two standard retrieval baselines: sparse retrieval (\textsc{sr}) using BM25, and dense retrieval (\textsc{dr}) using BGE-M3 embeddings~\citep{chen-etal-2024-m3}. Both rank candidates by cosine similarity without LLM capabilities or attribute extraction.

(ii) \noindent\textbf{\textsc{dr}+raw-ranker.} This baseline represents an LLM ranking approach without structured attributes. We retrieve the top 50 candidates using the same retrieval mechanism as above and then prompt Claude 3.5 Sonnet v2 to rank candidates based on their full product information (unstructured text).

Our proposed method is denoted as \textbf{\textsc{dr}+graph-ranker.} It enhances the ranking process by retrieving the top 50 candidates using the same retrieval mechanism, then replacing raw product text with structured attribute data from our product graph. Using an LLM to perform attribute-based comparison and ranking demonstrates the value of our structured attribute representation approach.
Given our zero-shot setting, we do not compare against cross-encoders or GNN based methods requiring task-specific training data. %
Our baseline design targets a controlled comparison: all systems share the same retrieval stage, and the two LLM rankers (raw-ranker vs. graph-ranker) use the same LLM, differing only on structured attribute graph input. Given the absence of comprehensive ground truth labels, we focus on precision-based metrics rather than recall, following established practices for top-k result evaluation \cite{Sakai2008}.

\noindent \textbf{Human Evaluation:}
We sample 40 diverse query product entities, yielding 200 unique query-candidate pairs per system (40 $\times$ 5 candidates). Three expert annotators assessed all four systems, producing 800 total annotated pairs. Annotators reviewed product information including titles, descriptions, and extracted attributes, providing binary judgments \texttt{<SIMILAR / NOT\_SIMILAR>} on whether a candidate can serve as a substitute for the query product. We obtained a high pairwise inter-annotator (Cohen's $\kappa$) agreement of 0.71.

Table~\ref{tab:human_eval} shows that our \textbf{\textsc{dr}+graph-ranker} method outperforms all baselines. Notably, precision@5 substantially improves from 39.52\% to 47.14\%, indicating that our method is particularly effective at retrieving relevant items deeper into the ranked list, thereby providing more comprehensive coverage of correct matches. We also observe consistent improvements in Mean Reciprocal Rank (MRR) from 60.00\% to 63.97\% and Mean Average Precision (mAP) from 57.99\% to 60.42\%, suggesting that our structured, graph-based ranking framework yields not only higher-precision top results but also a more robust overall ranking.
A qualitative error analysis reveals that our approach correctly matches products on specific attributes. E.g., for a query product entity titled ``\textit{kangaroo Home Security System | 5-Piece Kit | Compatible with Alexa and Google Home | App-Based | Pet-Friendly | Reduces Insurance Premium}'', our method correctly retrieves competing smart home security kits with matching component count, while the \textit{raw-ranker} tends to surface less relevant smart home devices with differing key attributes.

\noindent \textbf{LLM-based Evaluation:}
For automatic evaluation, we employ the Nova Pro LLM \cite{amazon2024nova}, a different LLM than the ranking model (Claude 3.5 Sonnet v2) to avoid self-evaluation bias. In a preliminary comparison on 500 examples, Claude 3.5 Sonnet v2 achieved 6\% higher precision@10 as a ranker but Nova Pro provided more conservative and discriminative relevance judgments. The evaluation LLM assesses query-candidate pairs based on attribute alignment and semantic similarity under a strict standard. 
Table \ref{tab:results} shows our system's performance (\textsc{dr}+graph-ranker) %
at two evaluation thresholds. %

At a higher similarity threshold ($\geq 80$), our method consistently outperforms all baselines. We observe substantial gains over the basic retrieval baseline: i.e., an $\text{nDCG}@1$ increase from 75.27\% (\textsc{dr}) to 79.96\% and a precision@1 gain from 49.08\% to 56.85\%. Moreover, the MRR score significantly improves from 55.57\% to 63.13\%, indicating that our proposed system both retrieves relevant entities and places them higher in the ranked results. While adding the \textsc{raw-ranker} component provides incremental improvements over the baseline retrieval, the substantial gap between \textsc{baseline-dr+raw-ranker} and our attribute-based ranking underscores the significant advantage of explicitly incorporating structured attribute information into ranking. 
At a more lenient threshold ($\geq 50$), our proposed method's precision@1 again rises from 74.78\% (\textsc{baseline-dr+raw-ranker}) to 78.12\%, and MRR increases  from 80.12\% to 83.53\%. These robust performance improvements across thresholds demonstrate the stable and meaningful benefits of structured attribute reasoning in entity similarity ranking tasks, effectively capturing nuanced similarities even under relaxed evaluation criteria.

We also separately evaluate the attribute extraction accuracy using Claude 3.5 Sonnet v2 as a judge on 20,000 samples, achieving 83.47\% F1. Hallucination is minimized through the predefined attribute schemas and explicit extraction instructions that constrain the LLM to find attributes in the provided text. %

\subsection{Live Production Deployment}
A first version of our proposed system is deployed in production at a leading global e-commerce service. The structured and standardized attribute schemas extracted from unstructured product information serve dual purposes: (i) as input to the graph-aware LLM ranking pipeline described above, and (ii) as training data for a high-quality in-house product embedding model. This embedding model captures nuanced product relationships, enabling retrieval of high-quality candidate products similar to the query product, which are then to be re-ranked via a product graph-aware LLM ranker, as described above. Additional business factors such as price and customer ratings can be incorporated as extracted attribute values into the attribute graph, and time-sensitive attributes (e.g., fashion trends) are captured through features like release year or technical specifications that naturally indicate recency. Existing structured data from the product catalog (e.g., brand fields) is directly incorporated into the attribute graph, with LLM extraction used only for attributes not already available in structured form.  

Our system serves as input for multiple downstream e-commerce applications, enabling additional business rules (e.g., pricing, availability) to be applied as a subsequent layer. 
Our embedding-based retrieval stage limits the online graph-aware LLM ranking component to a manageable candidate set (50--80 items). Structured attribute representations significantly reduce the context length fed to the LLM ranker -- approximately 300 tokens per product versus 700 for raw unstructured text descriptions, achieving a 57\% token reduction. For ranking 50 candidates, this reduces the total input by approximately 20,400 tokens, directly lowering attention complexity and KV cache memory requirements. In practice, this yields approximately 250 ms faster inference per request with Claude 3.5 Sonnet v2 compared to raw-text based ranking. While the online ranking component incurs inference costs relative to retrieval-only approaches, the latency remains feasible for high-value search scenarios such as entity similarity search in specialized product categories where precise attribute matching is critical for better quality outputs, user satisfaction and higher conversion rates. In-depth evaluation revealed metrics and trends matching or surpassing those reported above, with a preliminary human evaluation on 50 samples yielding a precision@3 of 0.76.

\section{Conclusion}
We present a two-stage system for entity similarity search that decouples LLM-driven attribute graph construction from graph-aware LLM ranking. By reasoning over structured attributes rather than raw text, our approach achieves a 6\% improvement in average precision over strong baselines in zero-shot settings, under practical considerations of scalability. Both human and LLM-based evaluations confirm consistent gains across metrics and product categories, with encouraging potential in a live deployment setting.

\clearpage

\section*{Presenter Biography}

Nikhita Vedula is a Senior Applied Scientist at Amazon. She received her Ph.D. from the Ohio State University. Her research interests and contributions span the fields of natural language processing, conversational AI and information retrieval. She has published in leading peer-reviewed conferences such as SIGIR, WSDM, the Web conference, EMNLP, NAACL and ACL. Her research has been recognized with one Best Paper award, one Best Paper Honorable Mention, and one Outstanding Paper award. She also regularly serves as a Program Committee member and Area Chair of multiple top tier conferences in her field. 

\bibliographystyle{ACM-Reference-Format}
\bibliography{anthology,custom}

\end{document}